\documentclass[11pt]{article}

\usepackage{amssymb,latexsym}
\usepackage[mathscr]{eucal}
\usepackage{citesort}

\newcommand{\bea}{\begin{eqnarray}}

\newcommand{\eea}{\end{eqnarray}}
\newcommand{\beq}{\begin{equation}}
\newcommand{\eeq}{\end{equation}}

\begin{document}
\title{A Note on Riemannian Anti-self-dual Einstein metrics with Symmetry}
\author{Paul Tod\\
Mathematical Institute and St John's College\\ Oxford}

\maketitle
\begin{abstract}
We give a complete proof of the result stated in \cite{tod1}, that
the general Einstein metric with a symmetry, an anti-self-dual Weyl
tensor and nonzero scalar curvature is determined by a solution of
the $SU(\infty)$-Toda field equation. We consider the two canonical
forms found for solutions to the same problem by Przanowski
\cite{prz1} and show that his Class A will reduce to the Toda
equation with respect to a second complex structure, different from
that in which the metric is first given.
\end{abstract}

\section{Introduction}
\label{SI}
In  \cite{tod1} it was claimed that the general
four-dimensional Riemannian Einstein metric with anti-self-dual (ASD) Weyl
curvature and a Killing vector could be written in a particular
form, based on a solution of the $SU(\infty)$-Toda field equation. A similar problem had been addressed by Przanowski \cite{prz1} a
little earlier. Having previously found that these metrics are necessarily Hermitian, he considered four-dimensional Hermitian-Einstein metrics with anti-self-dual (ASD) Weyl
curvature and a Killing vector, and he obtained two apparently different classes of
metrics, depending on the form of the Killing vector. One he
was able to
reduce to the $SU(\infty)$-Toda field equation, but the other he
was not.

In this note, we
shall give details of the proof of the result in \cite{tod1}, which have not
 previously appeared. This
provides an algorithm for reducing any particular ASD Einstein
metric with symmetry to the desired form. As such, it is close to
LeBrun's reduction of a scalar-flat K\"ahler metric with symmetry
\cite{L}, which it was inspired by. Applying the algorithm to a
metric in Przanowski's class B effects the reduction to Toda, which
he had already done. Applying it to a metric in his class A shows
that such a metric has a second complex structure, and it is with
respect to the second that the metric takes the desired form.

In the remainder of this Section, we shall review the construction
given in \cite{tod1}, and in the next we shall apply it to
Przanowski's two forms.

\medskip

Suppose then that we have a four-dimensional
Riemannian Einstein metric $g$ with ASD Weyl tensor and a Killing
vector $K$. We shall use the two-component spinor formalism \cite{PR}, adapted
for Riemannian signature. The derivative of the Killing vector
decomposes as
\beq
\nabla_aK_b=\phi_{AB}\epsilon_{A'B'}+\psi_{A'B'}\epsilon_{AB},
\label{K1}
\eeq
in terms of symmetric spinors $\phi_{AB}$ and $\psi_{A'B'}$. Any
Killing vector satisfies the identity
\[
\nabla_a\nabla_bK_c=R_{bcad}K^d\]
in terms of the Riemann tensor $R_{abcd}$, and with (\ref{K1}) and our
assumptions about the curvature this entails
\bea
\nabla_{AA'}\phi_{BC}&=&-\psi_{BCAD}K^D_{\;A'}+2\Lambda\epsilon_{A(B}K_{C)A'}\label{K2}\\
\nabla_{AA'}\psi_{B'C'}&=&2\Lambda\epsilon_{A'(B'}K_{C')A}\label{K3}
\eea
Here the conventions are slightly different from \cite{tod1}, to be in
line with \cite{PR}: $\psi_{ABCD}$ is the ASD Weyl spinor,
corresponding to the ASD Weyl tensor, and $\Lambda=R/24$ where $R$ is
the Ricci scalar.

Define the scalar $\psi$ by
\[2\psi^2=\psi^{A'B'}\psi_{A'B'},\]
and then define the tensor $J_a^{\;b}$ by
\beq
J_a^{\;b}=\psi^{-1}\delta_A^{\;B}\psi_{A'}^{\;B'}.
\label{j1}
\eeq
From the definition of $\psi$ it follows that $J_a^{\;b}$ is an almost-complex structure. Now from (\ref{K3}) we have
\[\nabla_{A(A'}\psi_{B'C')}=0\]
so that the spinor field $\psi_{B'C'}$ satisfies the {\emph{twistor equation}}
\cite{PR}. It then follows from a result of Pontecorvo \cite{MP} that
this almost-complex structure is integrable. (See the interpretation of this result in \cite{CP}; in our setting, it can be checked
directly: given (\ref{K3}) it is straightforward to calculate the
Nijenhuis tensor for $J_a^{\;b}$ and see that it vanishes.) Thus an
ASD Einstein metric with symmetry is necessarily Hermitian, as was
known to Przanowski, \cite{prz1}, \cite{prz2}, \cite{prz3}.

From (\ref{K3}) and the definition of the scalar $\psi$ we find
\[2\psi\nabla_{AA'}\psi=\psi^{B'C'}\nabla_{AA'}\psi_{B'C'}=2\Lambda\psi_{A'B'}K_A^{\;B'}\]
We introduce a coordinate $w=\Lambda\psi^{-1}$ and then the equation
above implies
\beq
J_{ab}K^adx^b=-\frac{dw}{w^2}.\label{K4}
\eeq
Guided by \cite{tod1}, we introduce a coordinate $\tau$ with
$K^a\partial_a\tau=1$ and a function $P$ related to the norm of the Killing vector by
\beq
Pw^2=(g_{ab}K^aK^b)^{-1}.
\label{K5}
\eeq
It follows from (\ref{K4}) and (\ref{K5}), since the complex structure preserves
lengths, that
\[g(\frac{dw}{w^2},\frac{dw}{w^2})=g(K,K)=(Pw^2)^{-1}.\]
Since the complex structure is integrable, we can introduce a complex
coordinate $\zeta=x+iy$ on the 2-plane orthogonal to
$K^a$ and $J_b^{\;a}K^b$. The metric now takes the form
\beq
g=\frac{P}{w^2}\left(e^v(dx^2+dy^2)+dw^2\right)+\frac{1}{Pw^2}(d\tau+\theta)^2.
\label{met1}
\eeq
for some function $v$ and one-form $\theta$. Note the similarity of
this to LeBrun's scalar-flat K\"ahler metrics \cite{L} (after
multiplication by $w^é$, it is precisely one of LeBrun's class of
metrics). To obtain an equation for $\theta$ we recall (\ref{K1})
and derive the pair of equations
\begin{eqnarray*}
K^b\nabla_aK_b&=&-\phi_{AB}K_{A'}^{\;B}-\psi_{A'B'}K_A^{\;B'}\\
K^b(\nabla_aK_b)^*&=&\phi_{AB}K_{A'}^{\;B}-\psi_{A'B'}K_A^{\;B'}
\end{eqnarray*}
where the star indicates Hodge dual. Using (\ref{j1}), (\ref{K4}) and (\ref{K5})
in these, we obtain
\[\frac{1}{2}dx^a\epsilon_{ab}^{\;\;\;\;cd}K^b\nabla_cK_d=\frac{dP}{2P^2w^2}+\frac{dw}{Pw^3}+2\Lambda\frac{dw}{w^3},\]
which translates into the following equation on $\theta$:
\beq
d\theta=P_{,x}dy\wedge dw+P_{,y}dw\wedge dx+e^v(P_{,w}+2P(1+2\Lambda
P)w^{-1}))dx\wedge dy.
\label{met5}
\eeq
This equation will have an integrability condition that we must return to.

\medskip

To make further progress, we need to impose the ASD Einstein
equations. A convenient route to these is as follows: suppose $\phi_i$
for $i=1,2,3$ is an orthonormal basis of SD 2-forms and introduce the
connection 1-forms $\alpha_{ij}=-\alpha_{ji}$ of the metric connection on these by
the equation
\[d\phi_i=-\alpha_{ij}\wedge\phi_j\;;\]
the ASD Einstein equations are now the system
\[d\alpha_{ij}+\alpha_{ik}\wedge\alpha_{kj}=2\Lambda\epsilon_{ijk}\phi_k.\]
For the metric (\ref{met1}), a convenient orthonormal tetrad is
\bea
\theta^1&=&w^{-1}P^{1/2}e^{v/2}dx,\label{t1}\\
\theta^2&=&w^{-1}P^{1/2}e^{v/2}dy,\label{t2}\\
\theta^3&=&w^{-1}P^{1/2}dw,\label{t3}\\
\theta^4&=&w^{-1}P^{-1/2}(d\tau+\theta),\label{t4}
\eea
and then the orthonormal basis of SD 2-forms can be taken to be
\bea
\phi_1&=&\theta^1\wedge\theta^4+\theta^2\wedge\theta^3,\label{f1}\\
\phi_2&=&\theta^2\wedge\theta^4+\theta^3\wedge\theta^1,\label{f2}\\
\phi_3&=&\theta^3\wedge\theta^4+\theta^1\wedge\theta^2.\label{f3}
\eea
For the connection one-forms we find
\begin{eqnarray*}
\alpha_{31}&=&A\theta^1,\\
\alpha_{23}&=&-A\theta^2,\\
\alpha_{12}&=&\frac{1}{2}(v_{,y}dx-v_{,x}dy)-\frac{2\Lambda}{w}(d\tau+\theta),
\end{eqnarray*}
where
\[A=-\frac{1}{2}P^{-1/2}(wv_{,w}-4\Lambda P-4).\]
Now impose the ASD Einstein equations to find first a condition
defining $P$:
\beq
-4\Lambda P=2-wv_{,w},\label{p1}
\eeq
and second an equation for $v$:
\beq
v_{,xx}+v_{,yy}+(e^v)_{,ww}=0,
\label{met2}
\eeq
which one recognises as the $SU(\infty)$-Toda field equation.

This is all that one gets from the ASD Einstein equations. With the
aid of (\ref{p1}), (\ref{met5}) simplifies to
\beq
d\theta=P_{,x}dy\wedge dw+P_{,y}dw\wedge dx+(Pe^v)_{,w}dx\wedge dy,
\label{met6}
\eeq
The integrability condition is
\[P_{,xx}+P_{,yy}+(Pe^v)_{,ww}=0,\]
which is easily seen to be satisfied by virtue of (\ref{p1}) and (\ref{met2}).

In summary, the metric is (\ref{met1}), where $v$ is chosen to be a
solution of the $SU(\infty)$-Toda field equation, $P$ is given by
(\ref{p1}) and $\theta$ is obtained by solving (\ref{met6}). The
freedom in the choice of $\theta$ corresponds to the freedom in the
choice of origin for $\tau$. Following the derivation of the metric in this form, we clearly have an algorithmic route to put any ASD Einstein metric with a symmetry into this form.

In the next section, we review Przanowski's form for ASD
Hermitian-Einstein metrics with symmetry and show how to reconcile
them with the form given here.

\section{Przanowski's forms for the metric}
\label{sPM}
Przanowski \cite{prz1} (see also \cite{prz2}, \cite{prz3}) shows that an ASD Einstein metric can locally
be written in terms of complex coordinates $Z^\alpha=(Z^1,Z^2)$ and a
single real function $h$ in the form
\beq
g=2g_{\alpha\overline{\beta}}dZ^\alpha dZ^{\overline{\beta}}\label{pm1}
\eeq
where, as usual, $Z^{\overline{\alpha}}=\overline{Z^{\alpha}}$, and
\beq
g_{\alpha\overline{\beta}}=-\frac{3}{l}\left(\partial_\alpha\partial_{\overline{\beta}}h+2\delta_\alpha^{\;2}\delta_{\overline{\beta}}^{\;\overline{2}}\;e^h\right),\label{pm2}
\eeq
where $l=6\Lambda$, in terms of $\Lambda$ from Section 1.

The ASD Einstein equation reduces to Przanowski's `master equation'
\cite{prz1}
\beq
\partial_1\partial_{\overline{1}}h\;\partial_2\partial_{\overline{2}}h-\partial_1\partial_{\overline{2}}h\;\partial_2\partial_{\overline{1}}h+(2\partial_1\partial_{\overline{1}}h-\partial_1
h\partial_{\overline{1}}h)\;e^h=0.\label{pm3}
\eeq
Now if $g$ has a Killing vector $K$, then Przanowski argues that it must
be possible to arrange the coordinates so that one of the following
two cases holds:

Class A:
\beq
K=i(\partial_2-\partial_{\overline{2}})\label{pm4}
\eeq

Class B:
\beq
K=i(\partial_1-\partial_{\overline{1}})\label{pm5}
\eeq
Let us attempt to reconcile the metric of Class B with what we have in
Section 1. This has been done already in \cite{prz1}, but it prepares
the way for Class A. Written out at length, the metric (\ref{pm1}) is
\[g=-\frac{6}{l}\left(h_{,1\overline{1}}\;dZ^1
dZ^{\overline{1}}+h_{,1\overline{2}}\;dZ^1
dZ^{\overline{2}}+h_{,2\overline{1}}\;dZ^2
dZ^{\overline{1}}+(h_{,2\overline{2}}+2e^h)\;dZ^2 dZ^{\overline{2}}\right),\]
with $h_{,1}=h_{,\overline{1}}$.

The function $P$ of (\ref{K5}) is obtained from the norm of the
Killing vector via the equation
\beq
P=w^{-2}(K^aK_a)^{-1}=-w^{-2}(\frac{6}{l}h_{,1\overline{1}})^{-1},\label{pm6}
\eeq
while the function $w$ is obtained from (\ref{K4}) by solving
\[\frac{dw}{w^2}=J_{ab}K^adx^b=-\frac{3}{l}(h_{,1\overline{1}}(dZ^1+dZ^{\overline{1}})+h_{,1\overline{2}}dZ^{\overline{2}}+h_{,2\overline{1}}dZ^2),\]
so that $w^{-1}=\frac{3}{l}h_{,1}$.

The Killing vector lowered with the metric is
\[K_adx^a=-\frac{3i}{l}(h_{,1\overline{1}}(dZ^{\overline{1}}-dZ^1)+h_{,1\overline{2}}dZ^{\overline{2}}-h_{,2\overline{1}}dZ^2.\]
With this, (\ref{pm6}) and the formula for $w$, we can compare with (\ref{met1}) to find
\beq
\frac{P}{w^2}e^v(dx^2+dy^2)=-\frac{6h_1h_{\overline{1}}}{lh_{1\overline{1}}}dZ^2dZ^{\overline{2}}
\label{pm7}
\eeq
so that the holomorphic coordinate $\zeta=x+iy$ can be taken to be $Z^2$, and then
\beq
e^v=4w^2e^h.
\label{pm8}
\eeq
It is now just a matter of checking that $v$ satisfies the Toda equation (\ref{met2}) by virtue of Przanowski's master equation (\ref{pm3}).

\medskip

If we try the same manipulations for Class A, with the Killing vector (\ref{pm4}), we may first calculate
\beq
J_{ab}K^adx^b=-\frac{3}{l}(h_{,1\overline{2}}dZ^1+h_{,\overline{1}2}dZ^{\overline{1}}+(h_{,2\overline{2}}+2e^h)(dZ^2+dZ^{\overline{2}})),\label{A1}
\eeq
but this is not (in general) exact, and so doesn't serve to define $w$. To see what has gone wrong, we calculate the self-dual part of the exterior derivative of $K$. By (\ref{K1}), this should be proportional to the complex structure. A lengthy but straightforward calculation leads to
\beq
(dK)^+=\frac{1}{2}h_{,2}\Phi+i\frac{h_{,1}}{|h_{,1}|}e^{h/2}\Psi-i\frac{h_{,{\overline{1}}}}{|h_{,1}|}e^{h/2}\overline{\Psi},
\label{A2}
\eeq
where
\begin{eqnarray*}
\Phi&=&-\frac{6i}{l}(h_{,1\overline{1}}\;dZ^1\wedge dZ^{\overline{1}}+h_{,1\overline{2}}\;dZ^1\wedge dZ^{\overline{2}}\\
&&\;\;\;\;\;\;\;\;\;+h_{,2\overline{1}}\;dZ^2\wedge dZ^{\overline{1}}+(h_{,2\overline{2}}+2e^h)\;dZ^2\wedge dZ^{\overline{2}})\\
\Psi&=&\sqrt{\Delta}\;dZ^1\wedge dZ^2,
\end{eqnarray*}
and
\[\Delta=\frac{36}{l^2}h_{,1}h_{,\overline{1}}\;e^h\]
which is (up to a constant) the determinant of the metric. $\Phi$ is the complex structure as a 2-form so that
\[\Phi\wedge\Phi=2\Psi\wedge\overline{\Psi}=-2\Delta dZ^1\wedge dZ^{\overline{1}}\wedge dZ^2\wedge dZ^{\overline{2}}.\]
Therefore the complex structure of Section 1, defined by the Killing vector, is different in Class A from the complex structure implicit in the metric written as in (\ref{pm1}). If we follow the algorithm of Section 1 with the appropriate complex structure, then we must as before reduce to the Toda equation. However it doesn't seem possible to find the holomorphic coordinates for the new complex structure explicitly - one simply has a proof of local existence.

One shouldn't be surprised to find a second integrable complex structure. For an ASD Einstein metric, there are locally many integrable complex structures, as follows from the Kerr Theorem of twistor theory in the Riemannian setting (see e.g. \cite{PR}): an integrable complex structure is determined by a geodesic shear-free spinor field and these are given by the vanishing of a (suitable, local) holomorphic function in twistor space. In deriving the metric form (\ref{pm1}), Przanowski \cite{prz3} begins from an arbitrary complex structure, though he needs to make a specific choice within the class of holomorphic coordinates. Then for his Class B, the complex structure derived from the Killing vector coincides with the one from which he starts, but for Class A it doesn't.

There is a connection with scalar-flat K\"ahler metrics which is
worth remarking: as noted above, the metric (\ref{met1}) is
evidently conformal to K\"ahler - multiply it by $w^2$ to obtain
LeBrun's form \cite{L} of the scalar-flat K\"ahler metric with
symmetry. Thus an ASD Einstein metric with a symmetry is conformal
to a scalar-flat K\"ahler metric. There is a partial converse of
this result based on the following result:

\medskip

{\emph{Suppose $(M,J,\,g)$ is a K\"ahler manifold, with complex
structure $J$ and metric $g$ and such that $(M,\tilde{g})$ is
Einstein with nonzero scalar curvature, where $\tilde{g}$ is
conformal to $g$, say $\tilde{g}=\Omega^2g$, then $g$ necessarily
has a symmetry.}

\medskip

This is given in \cite{AG} as part of a larger theorem for compact $M$ but the argument is purely local, and doesn't require compactness. Clearly if $g$ or $\tilde{g}$ is ASD, then both are and $\tilde{g}$ has the form (\ref{met1}).

To prove the statement in italics above, suppose the K\"ahler form can be written in spinors as $J_{ab}=J_{A'B'}\epsilon_{AB}$. The K\"ahler condition implies that
\[\nabla_{AA'}J_{B'C'}=0,\]
where $\nabla_{AA'}$ is the Levi-Civita covariant derivative for $g$. From the formulae for conformal rescaling (see e.g. \cite{PR}), it follows that
\beq
\tilde{\nabla}_{AA'}\tilde{J}_{B'C'}=2\tilde{\epsilon}_{A'(B'}K_{C')A}\label{j11}
\eeq
where $\tilde{J}_{B'C'}=\Omega^2J_{B'C'}$, $K_{AA'}=\Upsilon_{AB'}\tilde{\epsilon}^{B'C'}\tilde{J}_{A'C'}$ and $\Upsilon_{AA'}=\partial_{AA'}\log\Omega$. But now, by differentiating again and using the Ricci identities, the assumption that $(M,\tilde{g})$ is Einstein implies that
$$\tilde{\nabla}^{AB'}\tilde{\nabla}_A^{\;C'}\tilde{J}_{B'C'}=0=\tilde{\nabla}_{(B}^{\;A'}\tilde{\nabla}_{A)A'}\tilde{J}_{B'C'}$$
for any symmetric spinor $J_{B'C'}$. Substituting from (\ref{j11}) into this, we see that $K^a$ is a Killing vector.

\bigskip
\section*{Acknowledgements}
I acknowledge useful discussions with David Calderbank, Maciej
Dunajski and Nigel Hitchin. This note was prompted by a question
from Stefan Vandoren.

\end{document}